\documentclass[prb, aps,tightenlines,twocolumn,floatfix,superscriptaddress,showpacs,preprintnumbers,citeautoscript,10pt]{revtex4-1}
\usepackage{graphicx,verbatim,multirow,dcolumn}
\usepackage{mathtools}
\usepackage{amsmath,amssymb,bm}
\usepackage{comment}
\usepackage{mwe}
\usepackage{lettrine}
\usepackage{xcolor}
\usepackage{array}
\usepackage{float}
\usepackage{tikz}
\usepackage{url}
\usepackage{soul}
\usepackage[utf8]{inputenc}
\usepackage{lipsum}
\usepackage{natbib}
\usepackage{hyperref}

\makeatletter
\begin{document}
\title{Simulation of Atomic Layer Deposition with a Quantum Computer}

\author{
Evgeny Plekhanov
}
\affiliation{Quantinuum Ltd., 13-15 Hills Road, CB2 1NL, Cambridge, UK}
\author{
Georgia Prokopiou
}
\affiliation{Quantinuum Ltd., 13-15 Hills Road, CB2 1NL, Cambridge, UK}
\author{
Michal Krompiec
}
\affiliation{Quantinuum Ltd., 13-15 Hills Road, CB2 1NL, Cambridge, UK}
\author{
Viktor Radovic
}\affiliation{ C12, 26 rue des Fossés Saint-Jacques, 75005 Paris, France}
\author{
Pierre Desjardins
}\affiliation{ C12, 26 rue des Fossés Saint-Jacques, 75005 Paris, France}
\author{
Pluton Pullumbi 
}\affiliation{
Air Liquide, Paris Innovation Campus, 1 Chemin de la Porte des Loges, 78350 Les Loges en Josas, France}
\author{
David Mu\~{n}oz Ramo
}
\affiliation{Quantinuum Ltd., 13-15 Hills Road, CB2 1NL, Cambridge, UK}

\date{\today}

\begin{abstract}\label{section:abstract}
In this work, we present the study of 
an atomic layer deposition (ALD) of zirconium by means of a quantum computation on an emulator representing the features of an architecture based on qubits implemented on carbon nanotubes. 
ALD process control is key in several technological applications such as spintronics, catalysis and renewable energy storage.
We first derive a large ab-initio model of the precursor molecule approaching the infinite hydroxylated silicon (100) surface. In particular, we optimize geometry in three configurations: reactants, transition state and products.
Subsequently, we derive an effective small cluster model for each state. Atomic valence active space (AVAS) transformation is then performed on these small clusters, leading to an effective qubit Hamiltonian, which is solved using the Variational Quantum Eigensolver (VQE) algorithm.
We study the convergence of the reaction activation barrier with respect to the active space size and benchmark quantum calculations on a noiseless emulator and on an emulator representing a carbon nanotube qubit architecture, including an appropriate noise model and post-selection error mitigation. These calculations reveal an excellent agreement between the two emulation modes.
Our VQE calculations provide the multi-configurational corrections to the single determinant DFT and HF states and pave the way for the routine quantum calculations of ALD reactions.
\end{abstract}
\maketitle
    \section{Introduction}
The precise control of film thickness and the 3D conformability of the surface coatings by the Atomic Layer Deposition (ALD) has driven the development of materials with improved properties expanding the potential applications of ALD in several fields, such as magnetism and spintronics~\cite{quinard_ferromagnetic_2022,jussila_atomic_2023}, catalysis~\cite{sarnello_design_2021,knemeyer_mechanistic_2020}, bio-nanotechnology~\cite{bishal_atomic_2016,pessoa_chapter_2019}, photonics, clean and sustainable energy conversion~\cite{zhao_advanced_2021,elam_cheminform_2011} and renewable energy storage~\cite{zhu_surface_2023,riyanto_review_2021}, among others.
In particular, ALD technology has been and continues to be the key enabler for the continuous progresses in device microelectronics engineering~\cite{hwang_atomic_2014} following the trend of shrinking of component sizes. For instance, the state-of-the-art transistor technology depends entirely on ALD-grown materials with a large dielectric constant such as ZrO$_2$, requiring the use of a zirconium precursor in the ALD process together with a second reactant that is the source of oxygen. The surface chemistry involves two or more complementary and self-limiting steps in order to attain sub-monolayer control of the film growth~\cite{puurunen_surface_2005}.

The operating conditions of the ALD process and the quality of the deposited film strongly depend on the properties of the precursor, its reactivity with the substrate’s surface groups known as the first half-reaction as well as the reactivity of the oxygen source with the chemisorbed precursor on the surface known as the second half-reaction~\cite{zaera_mechanisms_2013}. 
In this context, the adoption of computational approaches for supporting the experimental investigations appears an obvious choice. 
Among the available computational methods, density functional theory (DFT)~\cite{mardirossian_thirty_2017} is the most commonly used due to the tradeoff between accuracy and computational cost in describing such complex systems. Several groups have used DFT-based methodologies and approximations~\cite{sibanda_review_2022} for investigating different aspects of the ALD process. However, there are still several characteristics of DFT such as the self-interaction error~\cite{bao_self-interaction_2018} that limit its accuracy for specific systems as the prediction of intermediates, transition states and activation barriers~\cite{kaplan_understanding_2023} in particular when dealing with transition metal organometallic complexes~\cite{shee_revealing_2021,semidalas_mobh35_2022} as is the case of precursors used in ALD deposition of ZrO$_2$ film. 

For such systems not accurately described by conventional computational methods and computers, quantum computers may provide an advantage~\cite{ivanov_quantum_2023,our_bmw_paper_2024}. This technology is expected to enable great speedups in the simulation of molecules and materials due to its efficient processing of the wavefunction describing a quantum system. 
Algorithms like Quantum Phase Estimation are being developed in order to exploit these features. Current quantum devices are still limited pushing the use of these very efficient algorithms to a later time. Nonetheless, a strong effort is being developed to test these machines for small proof-of-concept quantum chemistry simulations. These proofs of concept have been so far limited to small systems, or to large models where a small quantum region is defined via an embedding approach. In addition, it is common to perform these tests using algorithms requiring a much smaller circuit depth, like the Variational Quantum Eigensolver (VQE)~\cite{vqe2014} or quantum subspace methods~\cite{QSE_rev}.

Many different qubit architectures are being explored for the realization of this computing paradigm. Most common implementations include superconducting transmons, trapped ions or neutral atoms. More recently, a novel idea has been proposed: the use of electronic states on carbon nanotubes as qubits. This architecture has been investigated in \cite{SpinQCNTprop,HighCoherStateCNT,MicroSecCNT}. Physical realizations of this architecture are under development, but their performance may already be studied via the use of classical emulators where the main features of their operation have been coded, including their gate set and noise profile. These emulators enable the study of the performance of this technology in the execution of quantum algorithms for applications like quantum chemistry, which is the main subject of this paper.

In order to investigate these topics,
we consider here a typical chemical reaction involved in the ALD process as our test case. In particular, we focus on the first half-reaction taking place in the surface with ZyALD (tris(dimethylamino)cyclopentadienyl-Zirconium; [CpZr(N(CH$_3$)$_2$)$_3$]) as precursor and the hydroxylated Silicon (100) surface as substrate. We construct the models for the different species involved in this reaction, extract representative orbital active spaces, and perform energy calculations on a quantum emulator representing the operation of a quantum computer based on carbon nanotube qubits. The algorithm chosen for this analysis is VQE. This algorithm is a good choice for small quantum devices despite issues related to scaling of number of samples required and difficulty of the optimization process in large systems. These problems are not so important in small proofs of concept, which benefit from the small circuit depths required in VQE circuits. 

This paper is structured as follows. First, we will describe our methodology to create the models for the different species involved in our chosen reaction. Then, we will describe the quantum workflow used to perform the energy calculations of the different species, including details of the quantum backend selected. We will proceed then to present the results of our simulations. We will finish the paper with a discussion of the results and some conclusions.

    \section{Methods}

In this section, we describe the general workflow for our quantum simulations. We first optimize the geometries of the three structures using periodic boundary conditions in order to simulate the behavior of a large surface. For each one of the three states, we select a representative fragment from the optimized geometry. We perform HF total energy calculations in order to get an estimate of the relative energetics and the associated electronic structures. Based on these calculations, we perform fractional occupational density (FOD)\cite{Grimme2015} analyses in order to quantify and visualize the highly correlated regions of each structure. Subsequently, we use the Hartree-Fock (HF) results to construct an active space for each state, which we use for complete active space self-consistent field (CASSCF) calculations in order to recover the missing correlation at the HF level. This active space serves as a starting point for the upcoming calculations with a quantum computer. We feed the obtained molecular integrals and the selected active space into the VQE algorithm, and use as backend the Callisto quantum emulator. We describe these steps in more detail in the following subsections.

\subsection{Classical preprocessing}

As discussed in the literature~\cite{Ren2011,Jung2015,Hidayat2021,Rui2022,Van2022}, ALD processes include complicated surface reactions with several possible reaction paths with multiple transition states. In this work, we choose to focus on the first  part of the reaction, which involves the precursor approaching the surface (R), the state where the Zr atom is attached to one of the oxygens of the surface (P1) and an intermediate (TS) state, as presented in the Fig.~\ref{fig:QE_geom}. We follow closely the reaction path presented in Ref.~\onlinecite{Hidayat2021} for a Hf based precursor. 
Subsequently, we choose smaller fragments from the large periodic boundary conditions model and study on them the 
the accuracy of the complete active space self-consistent field (CASSCF) method~\cite{GKLC2017}, which is a special case of a more general multi-configuration self-consistent field (MCSCF) method~\cite{Helgakerbook}.

\subsection{Quantum workflow}

In this work, we employ the VQE method. VQE is based on the Rayliegh-Ritz variational principle and assumes a wave function of a definite functional form (ansatz) depending on a set of variational parameters, describing excitations over the Hartree-Fock (HF) state. 
VQE is a hybrid quantum-classical algorithm, where the expectation value of the Hamiltonian is optimized by the classical machine with respect to the ansatz's parameters set.
Here we use the unitary coupled cluster with singles and doubles excitations (UCCSD)~\cite{UCCSD} ansatz. In order to reduce the circuit depth of this ansatz, we use the adaptive derivative-assembled pseudo-Trotter (ADAPT) approach~\cite{adaptVQE}, as well as the so called chemically-aware circuit synthesis method~\cite{chemaware}. The ADAPT technique selects electronic excitations to be added to the ansatz circuit based on their derivatives with respect to the ansatz parameters. The chemically aware synthesis method reduces circuit depth further by analyzing the symmetry of the different excitations, removing forbidden ones, and mapping some double electron excitations to boson-like transitions.
At present, this combined framework can only solve relatively small chemical problems (without employing embedding techniques) in current noisy intermediate-scale quantum (NISQ) devices. Nevertheless, it is an appropriate tool to test and characterize quantum hardware.

\subsection{Quantum backend}
Several architectures are being proposed for qubit realization. In this paper, we explore qubits realized on carbon nanotubes via the Callisto emulator~\footnote{https://c12qe.github.io/c12-callisto-clients/, https://www.c12qe.com/callisto}. Callisto faithfully reproduces the behaviour of a quantum device based on the carbon nanotube architecture, including its noise profile.
 In this architecture, the qubit is encoded in the spin of a single electron hosted by a double quantum dot formed in a carbon nanotube~\cite{c12_2010,c12_2015,c12_2019}. The spin-spin interaction is mediated through virtual photon exchange with a microwave resonator. Various sources of noise have been considered in this architecture, the main ones being i) the Purcell effect coming from the resonator; ii) the noise due to charge fluctuations around the dots and iii) the phonon relaxation and de-phasing due to the mechanical motion of the tube. The noise coming from the phonons is peculiar for carbon nanotubes because the material is quasi one-dimensional. Four mechanical modes can be identified, from which only three contribute to the decoherence of the qubit (stretching, twisting and bending). The noise from all sources is taken into account by, first, determining the noise amplitude as a function of the qubits' parameters and then by determining the non-unitary evolution from Langevin and Lindblad type dynamics.
Finally, we have employed the partition-measurement symmetry verification (PMSV) error mitigation technique to discard the measurements that break fundamental system's symmetries like e.g. $Z_2$ mirror plane symmetry~\cite{PMSV}.

    \section{Classical modeling results}

\subsection{Large periodic model}

We first optimized the geometries of the R, TS, and P1 states with the standard Perdew-Burke-Ernzerhof (PBE) functional~\cite{Perdew1996} with the Van der Waals correction (Grimmme DFT-D3), as implemented in the \texttt{Quantum Espresso} code~\cite{QE2009, QE2017}.
We have used in-house generated Projector-Augmented Wave (PAW) sets and the plane wave cutoff of $60$ Ry. We have placed the system in a box with periodic boundary conditions (PBC) with $\Gamma$-point $k$-grid and a $10$ \AA $\;$thick vacuum along the $z$ direction. The resulting unit cell dimensions are: $15.397\times 15.397\times 40.444$ \AA.
As a substrate, we have used the hydroxylated Si slab with the $2\times 1$ reconstructed $(100)$ surface as defined in the Ref.~\onlinecite{Silicon}. The substrate,  hydroxylated on the top surface, was also passivated with hydrogen atoms on the bottom surface. The substrate was initially optimized separately, and subsequently the Si positions were maintained fixed, while the hydroxyl groups were allowed to relax during the deposition reaction. In total, in our calculations, the substrate contained as many as $80$ Si atoms for a total of $144$ Si, O, and H atoms.

\begin{figure*}
\centering
    \includegraphics[width=0.315\linewidth]{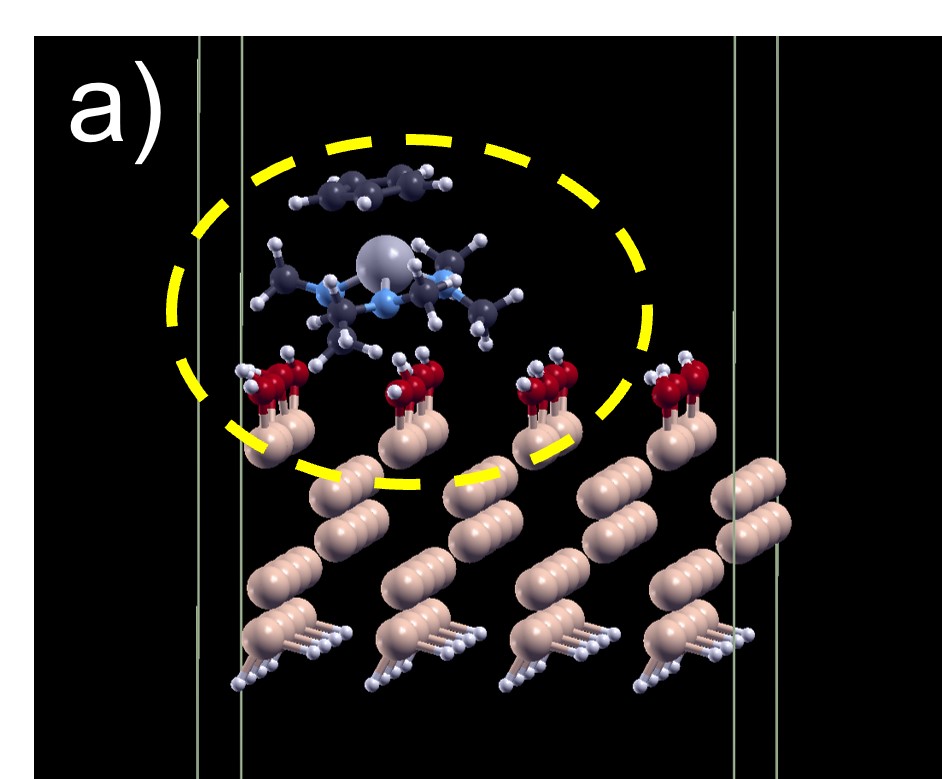}
    \includegraphics[width=0.34\linewidth]{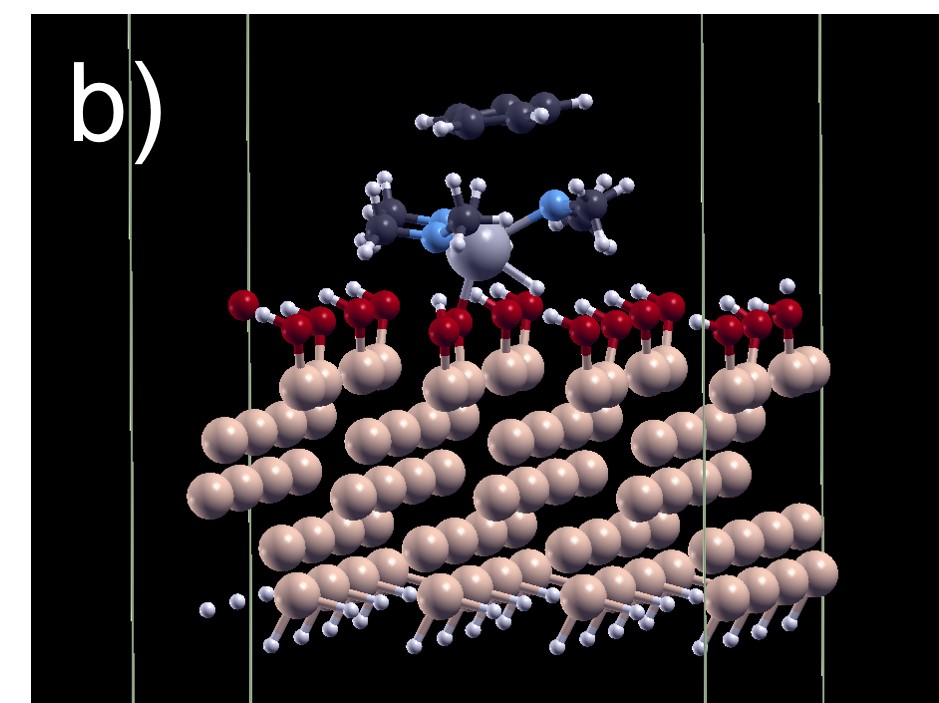}
    \includegraphics[trim=0 0.2cm 0 0.25cm,clip,width=0.32\linewidth]{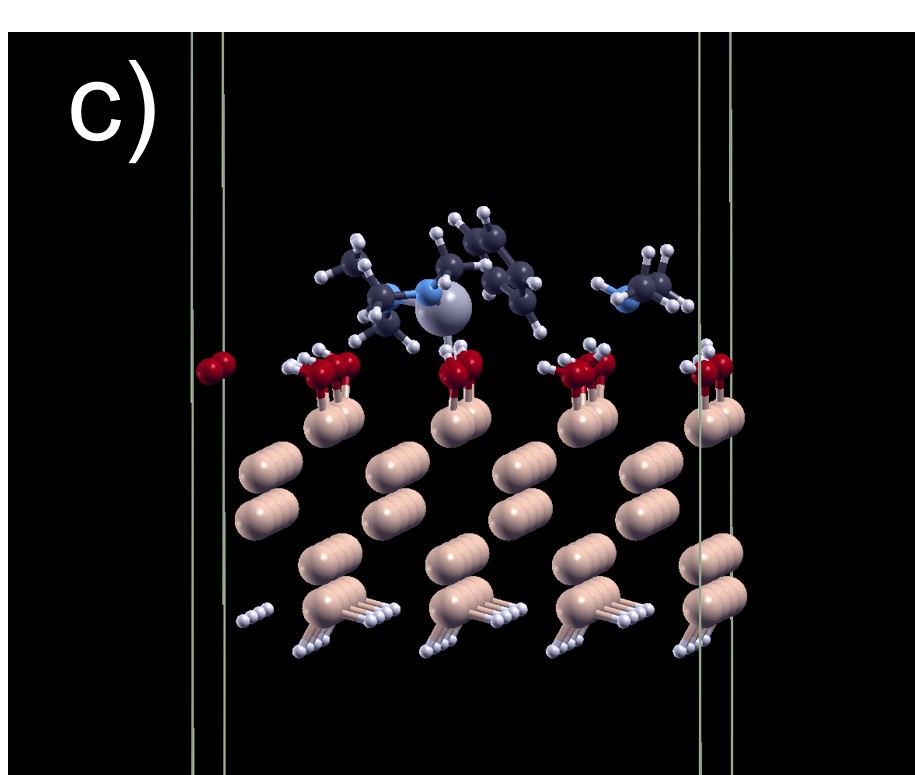}
\caption{Optimized geometries for \textbf{a)} the R state, \textbf{b)} the TS, and \textbf{c)} the P1 state. Colours of atoms: white: H, red: O, black: C, blue: N, gray: Zr, beige: Si. The solid lines represent the borders of the unit cell. The dashed yellow circle represents the fragment that is cut out of the structure. Images created using VESTA software package~\cite{VESTA}.
}
\label{fig:QE_geom}
\end{figure*}

The resulting optimized structures are shown in Fig.~\ref{fig:QE_geom}. In order to proceed with the calculations beyond DFT, we cut a smaller fragment out of the optimized structures, as depicted schematically in Fig.~\ref{fig:QE_geom}a.

    \subsection{Small clusters} 
\begin{figure*}[!h]
\centering
\begin{minipage}{0.47\textwidth}
\vspace{0pt}
    \includegraphics[width=1\textwidth]{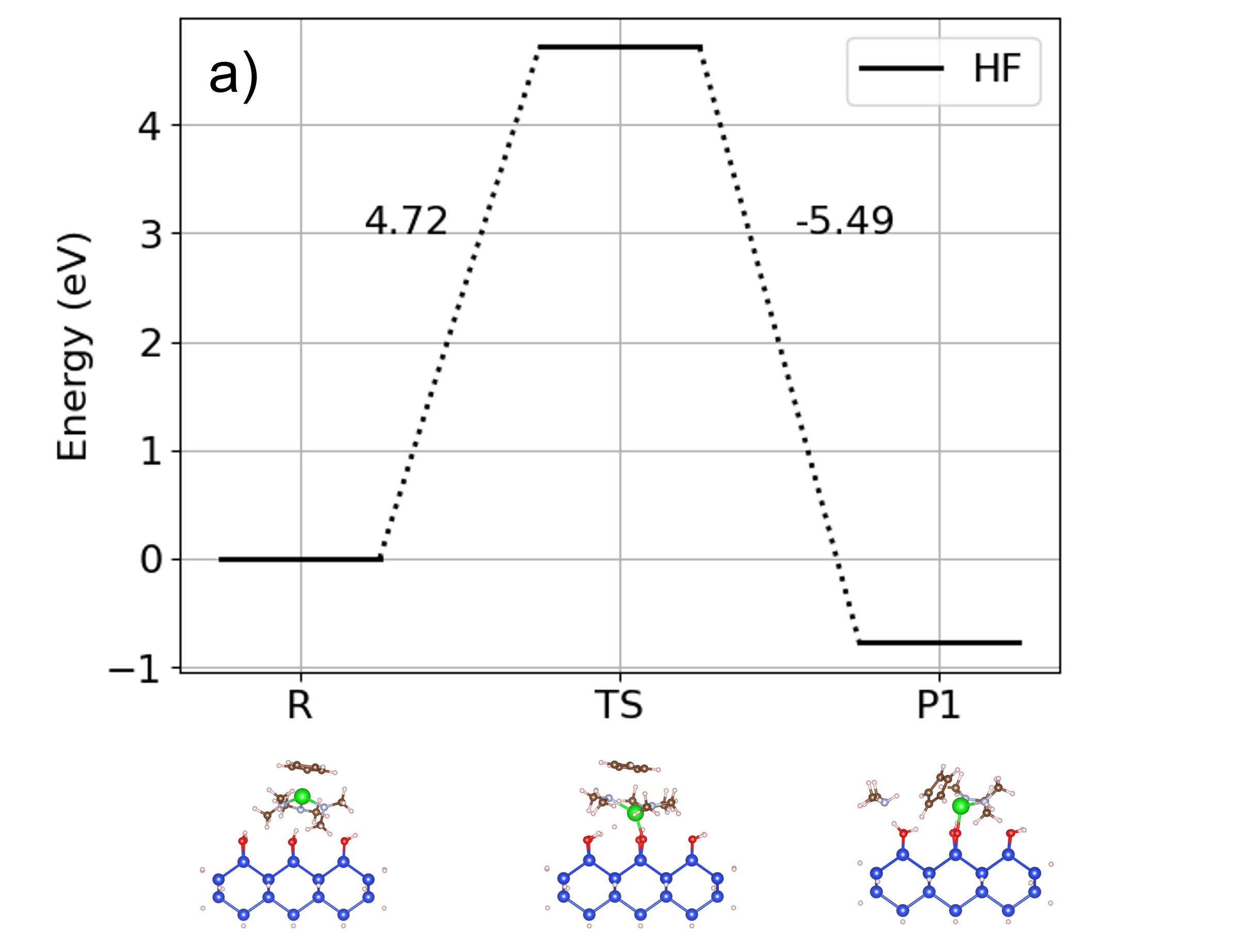}
\end{minipage}%
\begin{minipage}{.53\textwidth}
\vspace{0pt}
    \includegraphics[width=1.1\textwidth]{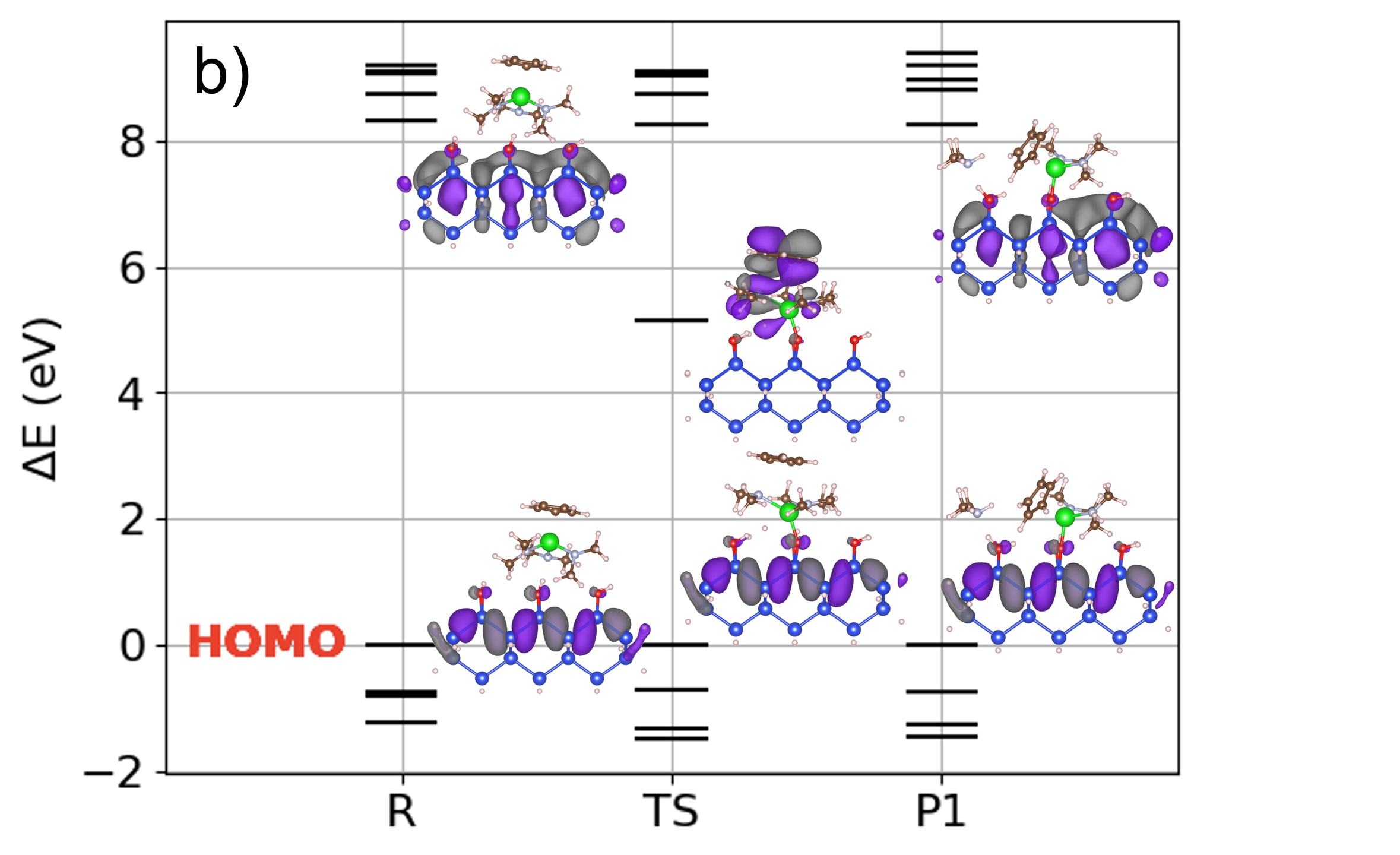}
\end{minipage}
\caption{
\textbf{a)} Relative energies of TS and P1 states with respect to the R state, obtained with HF. The gas-phase structures are shown in the bottom. DFT-D3 corrections were applied to the energy profile shown here. \textbf{b)} Valence eigenvalues with respect to the HOMO of each state for R, TS and P1 states. The HOMO and LUMO orbitals are also shown for each state. The isovalue was set to 0.02 and positive and negative regions are shown with purple and gray color, respectively. Colors of atoms: white: H, red: O, brown: C, cyan: N, green: Zr, blue: Si.
Images created using VESTA software package~\cite{VESTA}.
}
\label{fig:en_HF}
\end{figure*}

Subsequently, we have studied the electronic structure of the three states. We performed gas-phase calculations with the \texttt{PySCF}~\cite{pyscf} package. We used the def2-SVP basis set~\cite{Weigend2005} for all the atoms and the corresponding effective core potential (ECP) for the Zr atom. Grimme's dispersion corrections (DFT-D3) were also included~\cite{Grimme2010, Grimme2011}. We show in Fig.~\ \ref{fig:en_HF}a the HF total energy of R, TS and P1 states with respect to the total energy of the R state. 
It is well known that the single reference methods like DFT and HF do not describe reliably the compounds containing transition metal ions because of strong correlations  in partially filled $d$ shells~\cite{Truhlar2009}. As an indication of strong correlations, we consider the presence of $d$ orbital character in the valence states. The valence eigenvalues with respect to the highest occupied molecular orbital (HOMO) for the three states are shown in Fig.~\ref{fig:en_HF}b. The HF wave functions of the HOMO and the lowest unoccupied molecular orbital (LUMO) are also shown for each state, where positive and negative regions are plotted with purple and gray respectively. The alignment of the valence eigenvalues from the R to the P1 state remains relatively unchanged as most of the associated states are located on the surface (and the Si atoms more specifically) which does not undergo significant geometrical changes. However, in the case of the TS state, we observe a significant rearrangement of the energy levels. The LUMO of the TS state is dominated by the $d_{x^2-y^2}$ orbital of the Zr atom and the $\pi$ orbitals of the cyclopentadene, which indicate a higher degree of correlations.
Since HF highly overestimates the HOMO-LUMO gap~\cite{Kronik2008}, we only provide these results as a starting point for more accurate wave function based methods. 

\begin{figure*}[h]
    \includegraphics[width=0.33\linewidth]{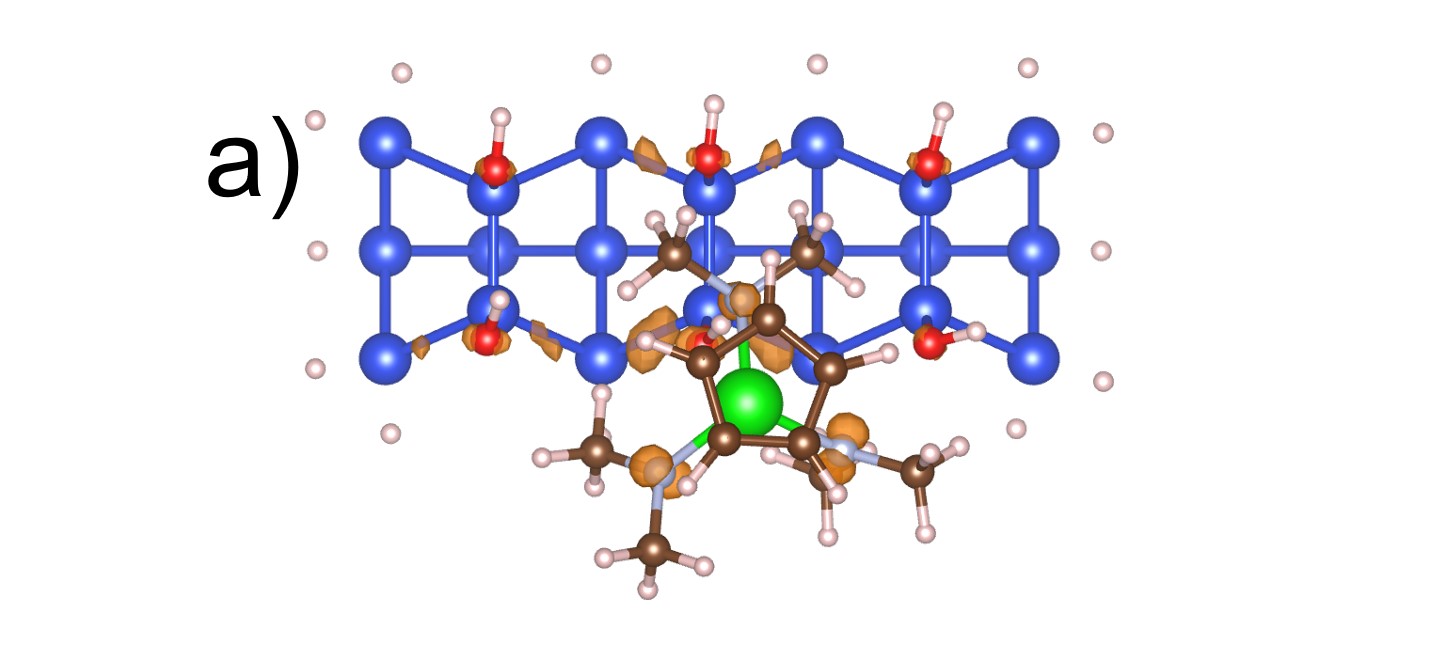}\hfill
    \includegraphics[width=0.33\linewidth]{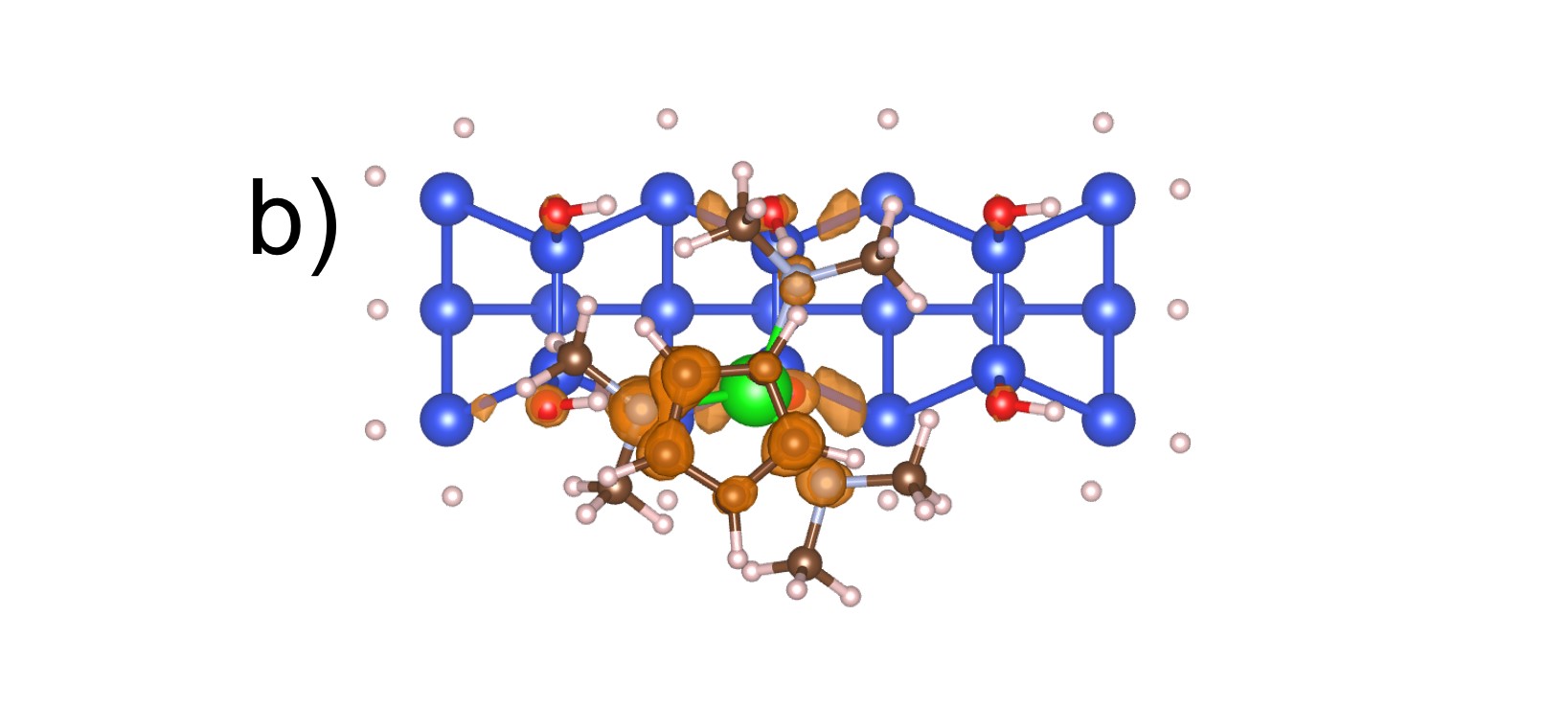}\hfill
    \includegraphics[width=0.33\linewidth]{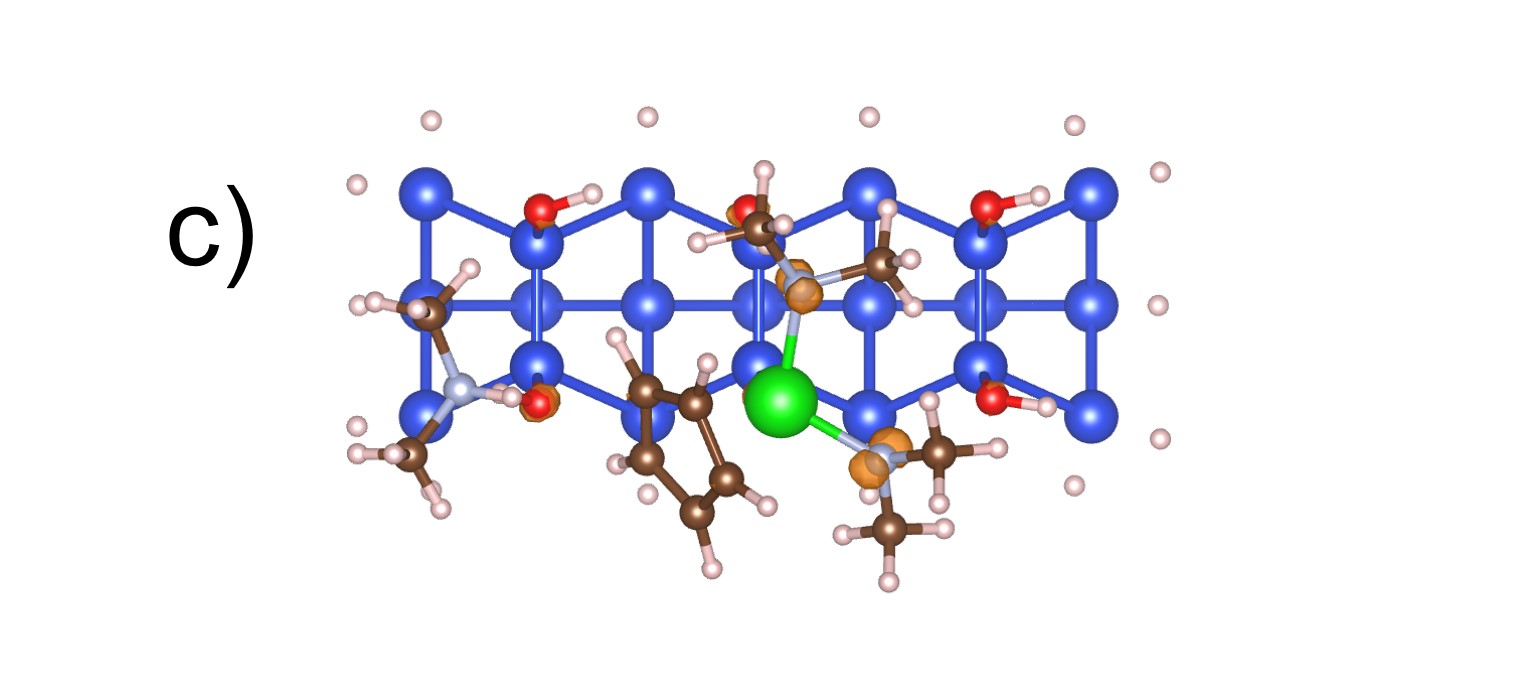}
\caption{
Fractional Orbital Density (FOD) plots for the \textbf{a)} R state, \textbf{b)} TS state, and \textbf{c)} P1 state. The isosurface is shown with orange and the isovalue was set to 0.005. The color index of the atoms is the same as in Fig.~\ref{fig:en_HF}.
Images created using VESTA software package~\cite{VESTA}.
}
\label{fig:fod_HF}
\end{figure*}

As mentioned above, the systems with strong static electronic correlations (SEC) are challenging cases for DFT, which makes them good test playground for quantum computers. A measure of SEC that was suggested recently\cite{Grimme2015} is the fractional orbital density (FOD). The FOD is constructed with fractional occupation numbers, which are calculated from DFT by using a finite-temperature Fermi-Dirac smearing for the electronic occupancies. As argued in the original paper, transition states usually present higher degree of correlations with respect to both reactants and products states. We present the FOD plots in Fig.~\ref{fig:fod_HF}, where the density is shown in orange. As expected, the TS state exhibits a stronger degree of SEC and the shape of the associated FOD is consistent with the analysis presented in Fig.\ \ref{fig:en_HF}.

    \subsection{AVAS, CASSCF} As discussed above, HF calculations provide only a qualitative picture of the energetics and the electronic structure in the case of a strongly correlated system. We have used the restricted HF (RHF) results as an initial guess to more accurate and more computationally expensive multiconfiguration (MC) methods. In order to reduce the size of the problem, MC methods require choosing \textit{a priori} an active space of molecular orbitals. To this end, we have employed the atomic valence active space (AVAS) method~\cite{avas2017}. This method is based on a linear transformation of occupied and unoccupied orbitals based on projectors of atomic valence orbitals, which are provided by the user. Here, we used the FOD of the TS state (see Fig.\ \ref{fig:fod_HF}b) in order to identify the most important spatial regions of the system. In addition, we chose the atomic orbital projectors based on the leading atomic orbital contributions to the valence molecular orbitals of the TS state (see Fig.\ \ref{fig:en_HF}b). We have used the $2p$ orbitals of the N and C atoms of the cyclopentanyl and the three O atoms closest to the Zr complex. We have also included the $s$ orbitals of the dissociating H atom and the $4d$ orbitals of the Zr atom, which are expected to be fractionally occupied in the TS state.

\begin{figure*}[h]
    \includegraphics[width=0.33\linewidth]{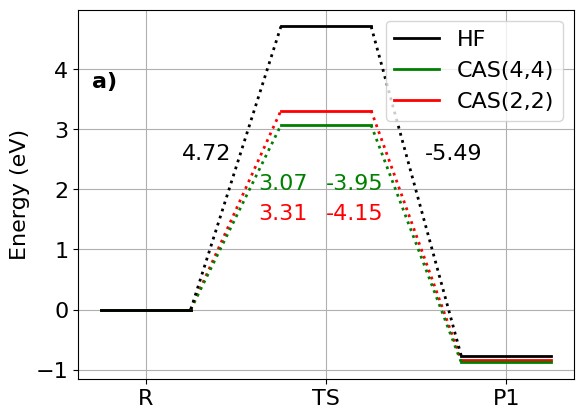}\hfill
    \includegraphics[width=0.32\linewidth]{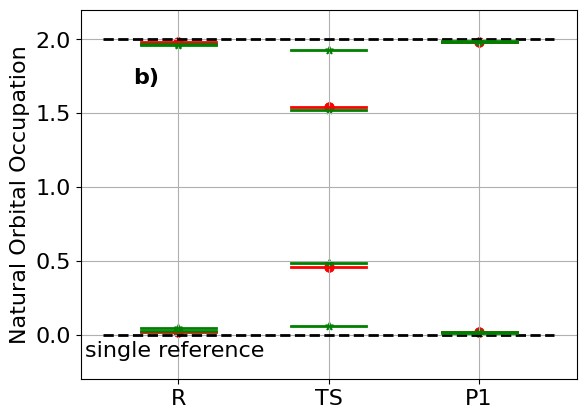}\hfill
    \includegraphics[width=0.34\linewidth]{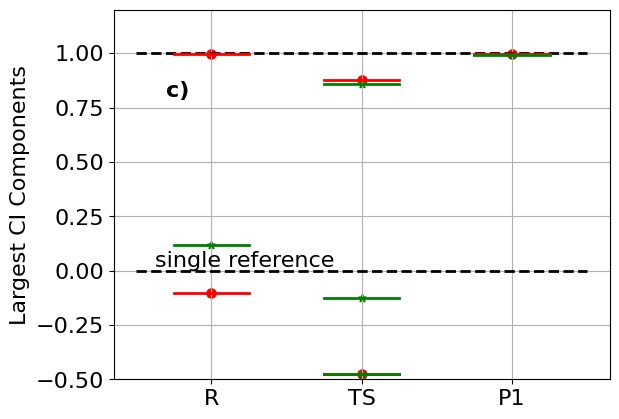}
\caption{\small{\textbf{a)} Relative energies of TS and P1 states with respect to the R state, obtained with HF (black). The CASSCF results with two active spaces i) of $4$ electrons and $4$ orbitals (green) and ii) $2$ electrons and $2$ orbitals (red) are also shown. DFT-D3 corrections were applied to the energy profile shown here.
\textbf{b)} Natural orbital occupation numbers and \textbf{c)} Largest CI coefficients obtained with CASSCF calculations. The colour indexes for all the graphs are indicated in the first panel. In the latter two graphs, black dashed lines represent the values of a single-reference system.}}
\label{fig:CAS}
\end{figure*}

The size of the resulting AVAS was still prohibitive ($41$ orbitals) in the context of quantum computers. In order to improve the energetics upon the HF calculation shown in Fig.~\ref{fig:en_HF}a we have performed CASSCF calculations with reduced active spaces based on the larger AVAS. The results are shown in Fig.~\ref{fig:CAS}a, where we observed that a minimal active space of two electrons in two spatial orbitals $(2e, 2o)$ is in good agreement with the results of a larger space $(4e, 4o)$. Furthermore, the relative energy difference of R and P1 states is hardly changed, while the TS state is more affected. This is because the R and P1 states are essentially single referenced states, as is shown in Fig.~\ref{fig:CAS}b and Fig.~\ref{fig:CAS}c. On the contrary, the occupation numbers of the orbitals of the TS state deviate from the values of a single-referenced system significantly (see \ref{fig:CAS}b) and the same is true for the largest CI coefficients (see \ref{fig:CAS}b), which for a single-referenced system would be $1.0$ and $0.0$. This analysis is in line with the qualitative analysis of the highly correlated (over the R and P1 state) character of the TS state, given in the previous section. 

    \section{Quantum Computing Results}

\subsection{Calculations on a noiseless quantum computer emulator}
Next, we characterize the performance of our quantum workflow in simulating the energetics of the reaction under study by repeating the CASSCF calculations shown in Fig.~\ref{fig:CAS} with a noiseless quantum emulator. More specifically, we map the CASSCF fermionic hamiltonian to a qubit hamiltonian ($\hat{H}$) using the Jordan-Wigner (JW) mapping. We also employ the chemically aware ansatz ($U(\theta)$) which has been shown to yield more compact circuits compared to other ansatze without compromising the accuracy~\cite{ChemAwareCQ}. We provide information on the generated ansatz circuits, before optimization, in table \ref{table_cas}.
\begin{table}
\begin{center}
\resizebox{\columnwidth}{!}{
\begin{tabular}{|c| c c c c|} 
 \hline
 & no. Qubits & circuit depth & circuit gates & no. parameters \\ [0.5ex] 
 \hline\hline
 CAS(2,2) & 4 & 8 & 17 & 1 \\ 
 \hline
 CAS(4,4) & 8 & 360 & 525 & 18 \\
 \hline
\end{tabular}

}
\caption{Quantum resources required for the circuits generated to run VQE with CAS(2, 2) and CAS(4, 4) active spaces.}
\label{table_cas}
\end{center}
\end{table}

\begin{figure}
    \includegraphics[width=\linewidth]{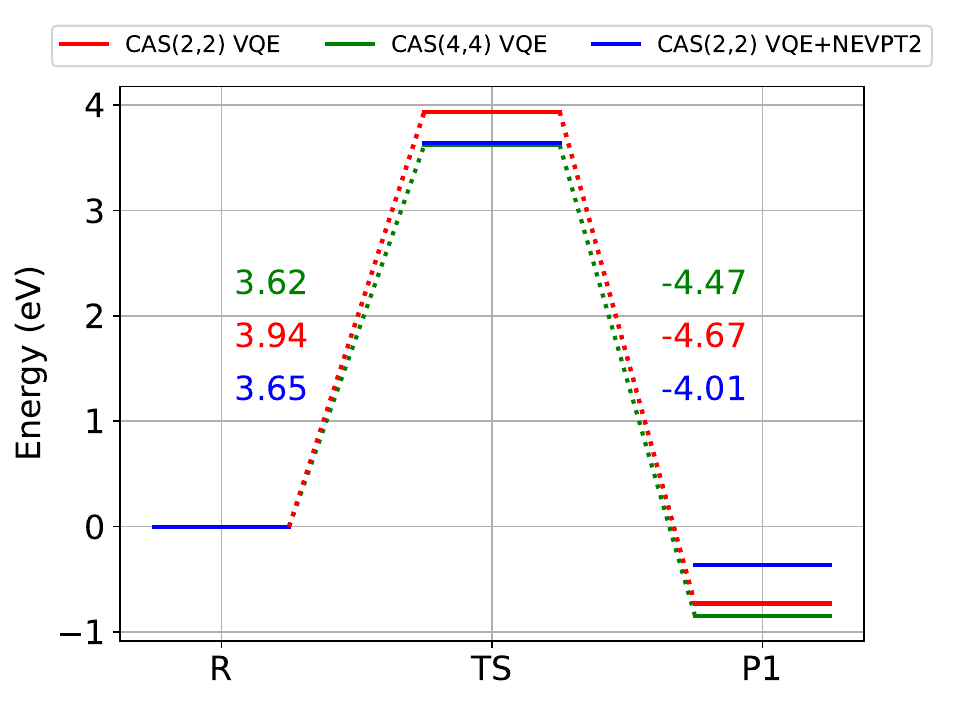}%\hfill
\caption{Relative energies of TS and P1 states with respect to the R state, obtained with VQE on the CASSCF orbitals with two active spaces: i) of $2$ electrons and $2$ orbitals (red), ii) $4$ electrons and $4$ orbitals (green), and iii) NEVPT2-corrected results for the $(2,2)$ active space (blue). DFT-D3 corrections were \textit{not} applied to the energy profile shown here.}
\label{fig:vqe}
\end{figure}

We obtain the minimum energy, 
$$
E = \left<\Psi_{\mathrm{HF}}|U^\dagger(\theta)\hat{H}U(\theta)|\Psi_{\mathrm{HF}}\right>$$
by varying the parameters within VQE\cite{vqe2014,InQuanto} and we show the results in Fig.~\ref{fig:vqe}. We note here that van der Waals corrections are not included in Fig.~\ref{fig:vqe} (\textit{cf} Fig.~\ref{fig:CAS}a); hence the differences between the two figures. Furthermore, we also calculate the n-electron valence state perturbation theory (NEVPT2) correction on the CASSCF results in order to improve the treatment of dynamic correlation. The results are shown in Fig.~\ref{fig:vqe} for the CAS(2,2) and they agree qualitatively with the respective CASSCF results in Fig.~\ref{fig:CAS}a.

    \subsection{Calculations on a quantum computer emulator with a noise model}

In this section, we show the results of quantum computations using the Callisto emulator as a backend, which includes a noise model corresponding to the carbon nanotube architecture applied in this work. We considered the (2, 2) active space for these simulations given Callisto emulation constraints.
The optimized variational gate-angles were obtained previously from the simulation on the noiseless emulator. They are kept fixed throughout the experiments. In Table \ref{callisto_tab}, we provide information on the shortest/longest circuit depth and number of gates of the ansatz circuits. These results are obtained after the transpilation phase, where we converted the original ansatz circuit to a circuit with a Callisto native gate set ($\mathrm{R}_x,\mathrm{R}_y, \mathrm{R}_z, i\mathrm{SWAP}$). This transpilation phase is performed within the IBM Qiskit framework \footnote{https://docs.quantum.ibm.com/transpile}.

\begin{table} 
\begin{center}
\resizebox{\columnwidth}{!}{
\begin{tabular}{|c| c c c c|} 
 \hline
 & no. Qubits & circuit depth & no. gates & \\ [0.5ex] 
  \hline\hline
Before transpilation & 4 & 29 & 42 & \\ 
 \hline
After transpilation & 4 & 7-12 & 10-16 &\\ 
 \hline
\end{tabular}
}
\end{center}
\caption{Quantum resources for the circuits used in CAS $(2,2)$ before and after transpilation to the Callisto native gate set.}
\label{callisto_tab}
\end{table}

\begin{figure}%[h]
    \includegraphics[width=1\linewidth]{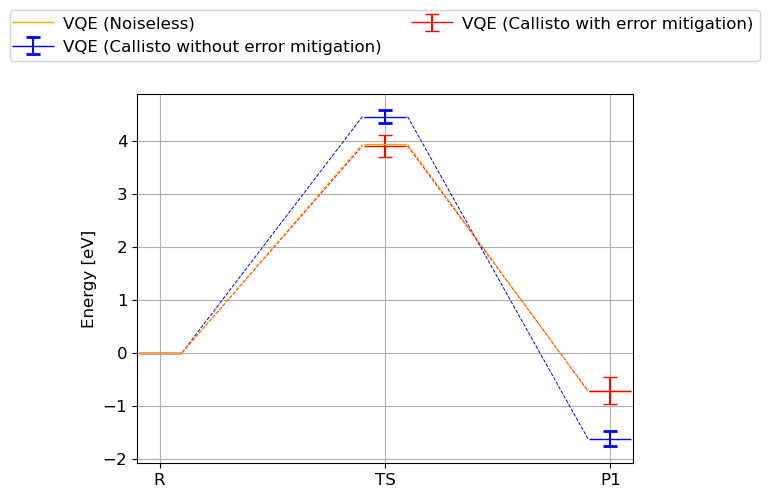}\hfill
\caption{\small{Relative energies of TS and P1 states with respect to the R state, obtained with VQE on Callisto without (blue) and with (red) error mitigation for two active space of $2$ electrons and $2$ orbitals. Relative energies obtained from VQE on a noiseless emulator (orange) are also shown for comparison.}}
\label{fig:callisto}
\end{figure}

We schedule $100$ experiments, each of them being the average of 100,000 shots. Fig.~\ref{fig:callisto} shows the noisy emulator results of these experiments. The energies of the TS and P1 states are shown relative to the energy of the R state, along with the results from a noiseless simulation. 

In the case of noisy simulations, the errors in estimating these energies have two components. One is the statistical uncertainty caused by the limited number of quantum measurements taken, which we represent by the standard deviations. The other contribution is due to the accumulation of operation errors during the quantum computation, which causes the final-state density matrix of the quantum computer to be a mixture of ground and higher excited states of the simulated system. We find that the incorporation of noise leads to a significant underestimation of the energy values with respect to the state vector simulation values in the three species considered. In the specific qubit architecture that we emulate, the decoherence channels are dominated by relaxation, in particular photonic dissipation, leading to an increase of ground states in the final distribution. This divergence does not exceed 300 mHartree thanks to the relatively shallow circuit depth for such a small active space, which prevents the circuit from accumulating large amounts of errors during operation. However, not all species are affected identically. The absolute energy bar of the TS state deviates less from the theoretical value than that of R, while the absolute energy bar of the P1 state deviates more than that of R. These differences increase the relative energy bar of the TS state, while pushing the P1 relative energy below the theoretical value.

We have also used the PMSV error mitigation using the $Z_2$ symmetry~\cite{PMSV} to improve the accuracy of the estimated energies. This post-processing eliminates the results forbidden by system's symmetries. Overall, the error mitigation makes all the three states get closer in energy to each other, which produces an increase of the P1 energy relative to R and, on the contrary, a decrease of TS's relative energy in comparison to R. As shown in Fig.~\ref{fig:callisto}, the results of noisy simulations followed by error mitigation are in excellent agreement with the results of noiseless experiments for the active space of $2$ electrons in $2$ orbitals. It is worth pointing out, however, that the application of the noise mitigation scheme increases the dispersion of values, hence the larger standard deviation values obtained for the noise-mitigated energies. This is a usual drawback in postselection noise mitigation.

Two conclusions can be drawn from these NISQ emulator calculations.
First, the energies predicted by the noisy simulations and their standard deviations stabilize around $10^5$ shots (see Fig \ref{fig:conv_sim} in the appendix), indicating that the quantum algorithm has converged. This finding suggests that the energy value found for $10^5$ shots is reproducible and will be found equally if the algorithm is re-run.
Second, the quantum workflow estimates the average energy values close to chemical accuracy (considered to be 1.6 mHartree) and manages to capture the correlation energy not included in the Hartree–Fock method. The application of a noise mitigation step is a key ingredient to obtain accurate results.

    \section{Discussion and Conclusions}
Here, we have presented a protocol for studying a reaction present in ALD with the complete active space self-consistent field (CASSCF) method, followed by the quantum computer calculations using the VQE approach on an emulated novel architecture based on carbon nanotubes.

The results obtained demonstrate that the VQE algorithm can describe the energetics of the ALD reaction considered.
Once noise coming from the Callisto emulator is added, significant errors are observed in the calculated VQE energies. These errors are successfully mitigated via the application of postselection noise mitigation. These results indicate that noise mitigation is an important ingredient to incorporate into a quantum workflow in the CNT architecture. Other noise mitigation techniques like zero noise extrapolation or probabilistic error cancellation may also be useful to improve results obtained from the quantum processor. 

Further work could explore how to improve the robustness of variational quantum algorithms to qubit errors, like the optimization of variational gate-angles. Further studies about dominant noise channels and tailored noise mitigation will also be needed to further exploit the features of the architecture considered in this work. At longer timescales, additional work both on hardware and algorithms will be required to scale this workflow to devices with larger numbers of qubits.

In conclusion, this works shows promising results to tackle an industry-relevant use case, the atomic layer deposition (ALD) of zirconium, by means of quantum computers and shows that a novel quantum computer architecture based on carbon nanotubes is appropriate for this task. Future improvements of the performance of quantum computers and novel quantum algorithms would be needed to deliver the full power of quantum computing.

    \section{Acknowledgments}

We would like to thank Q. Schaeverbeke, G. Christopoulou and C. Gaggioli for carefully reviewing the manuscript and contributing with ideas.
    \appendix
    \section{Convergence Study} 
\begin{figure*}[h]
    \includegraphics[width=0.50\linewidth]{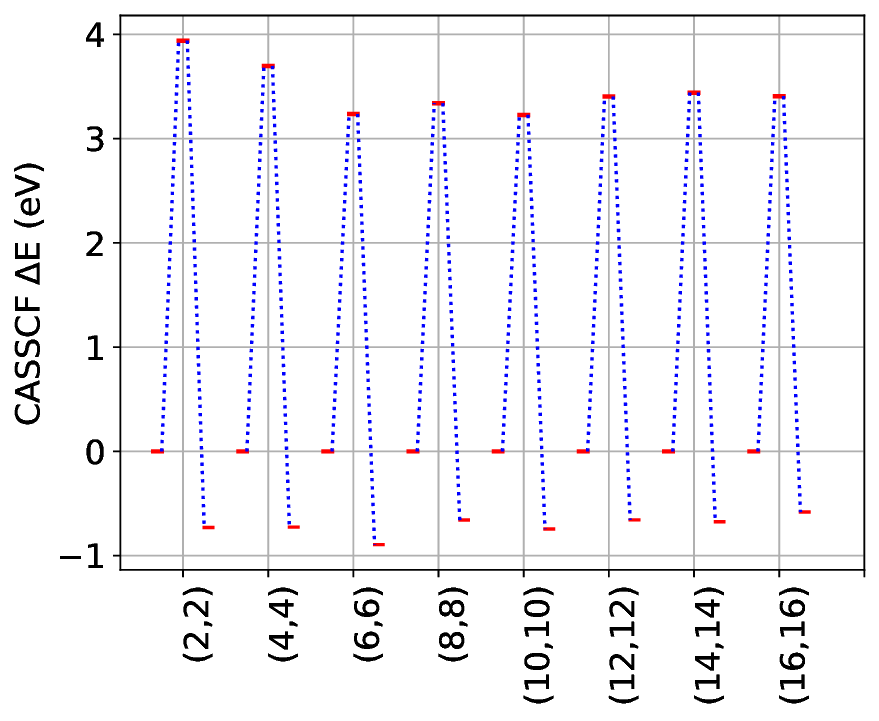}\hfill
    \includegraphics[width=0.50\linewidth]{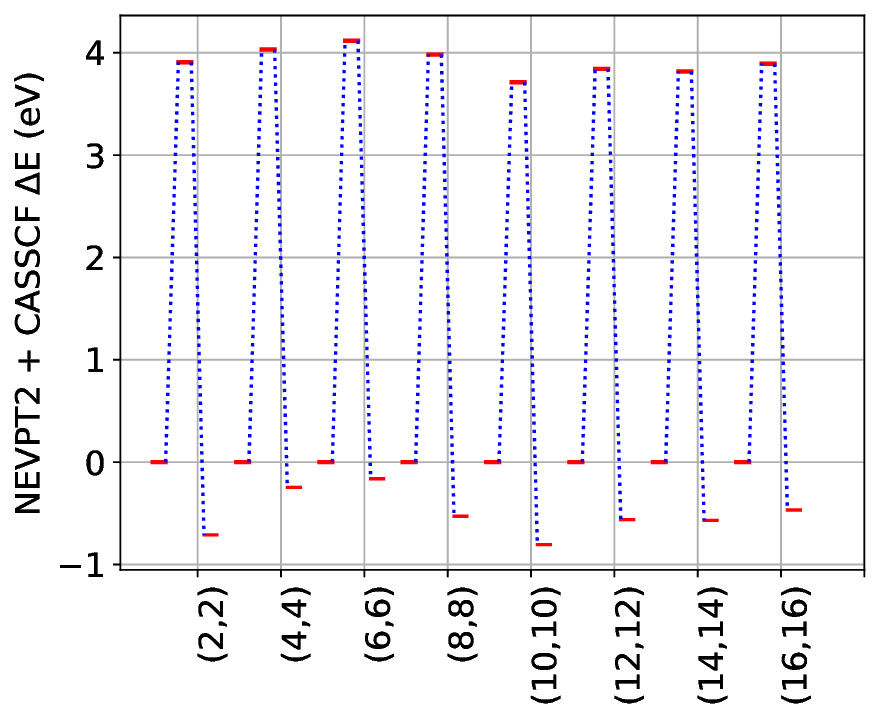}
\caption{\small{Left panel: relative energies of TS and P1 states with respect to the R state, obtained with CASSCF. The dimensions of the active spaces for each step is shown in the x-axis. Right panel: The same plot for the CASSCF energies corrected with the NEVPT2 values.}}
\label{fig:conv}
\end{figure*}

\begin{figure*}[h]
    \includegraphics[width=0.50\linewidth]{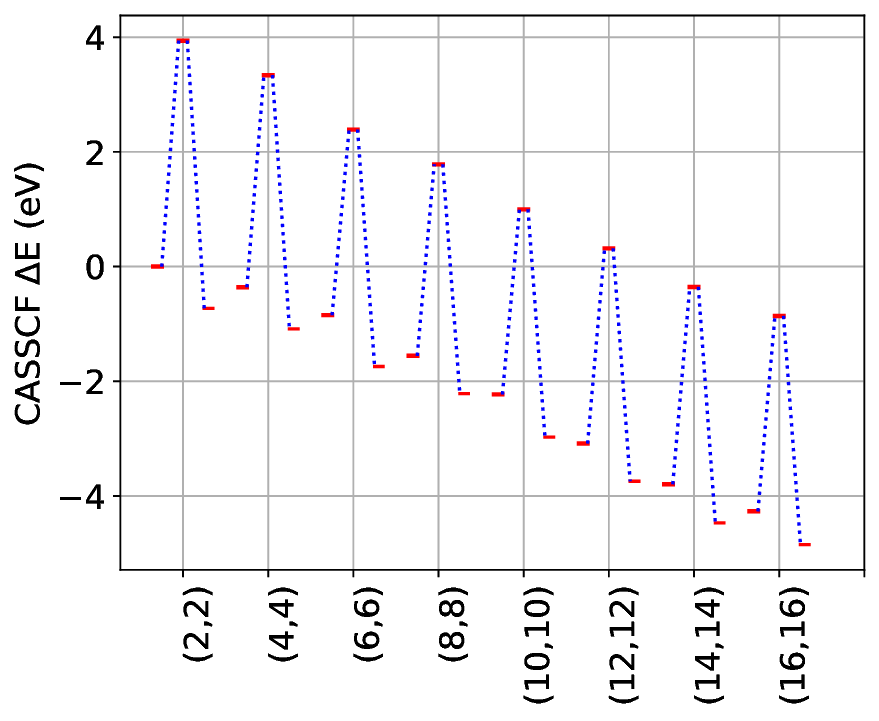}\hfill
    \includegraphics[width=0.50\linewidth]{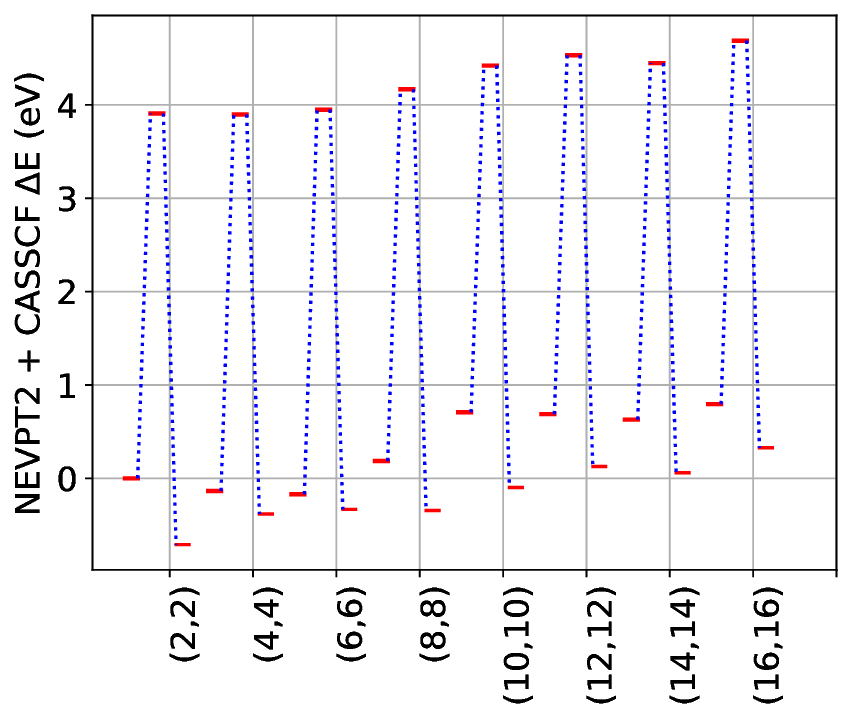}
\caption{\small{Left panel: relative energies of TS and P1 states with respect to the R state, obtained with a CAS(2, 2) calculation. The dimensions of the active spaces for each step is shown in the x-axis. Right panel: The same plot for the CAS(2, 2) energies corrected with the NEVPT2 values.}}
\label{fig:conv2}
\end{figure*}

Additionally, we have conducted a  convergence study of the CASSCF energy with respect to the size of the active space. The CASSCF energies for R, TS and P1 states at $(2,2)$, $(4,4)$, $(6,6)$, $(8,8)$, $(10,10)$, $(12,12)$, $(14,14)$, and $(16,16)$ active space sizes are shown in the left panel of Fig.~\ref{fig:conv} and Fig.~\ref{fig:conv2}. 
The TS and P1 state energies, relative to the R state one, tend to stabilize for large active spaces.
The qualitative trend of the energetics of the reaction remains unchanged as the size of the active space is increased.
A similar behavior is observed also for the CASSCF energies, when the NEVPT2 corrections are taken into account, as shown in the right panel of Fig.~\ref{fig:conv} and Fig.~\ref{fig:conv2}.
These findings demonstrate the robustness of our results with respect to the active space size and justify the use of the smaller active spaces throughout the project.
    \section{Convergence study of the simulations with a noise model}

\begin{figure*}[h]
    \includegraphics[width=0.32\linewidth]{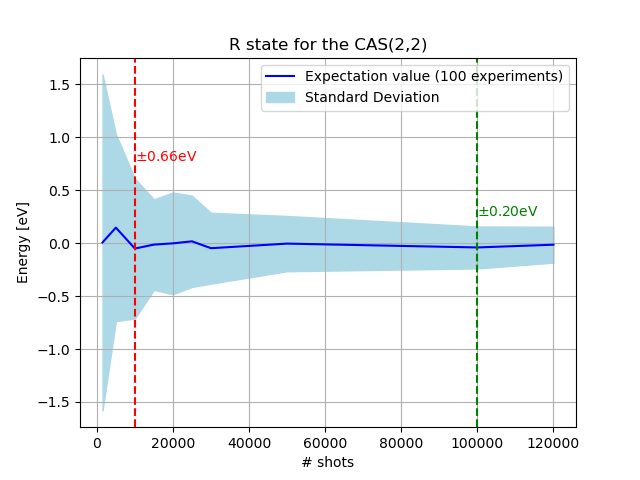}\hfill
    \includegraphics[width=0.33\linewidth]{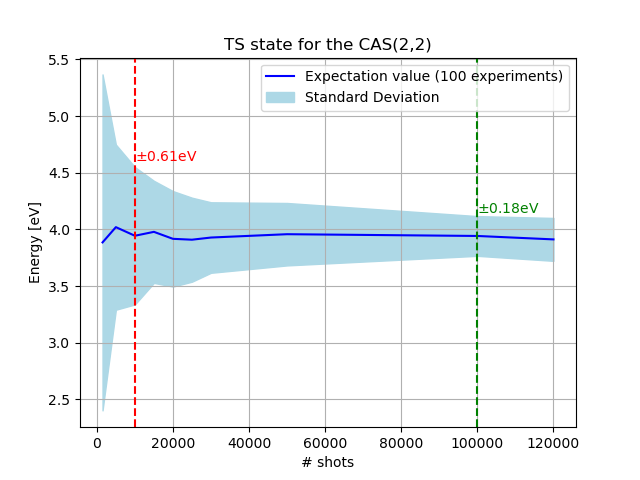}\hfill
    \includegraphics[width=0.34\linewidth]{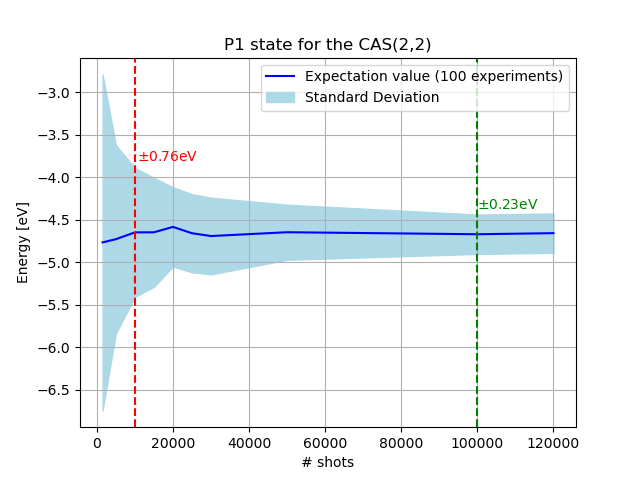}\hfill
\caption{\small{Absolute energies of R, TS and P1 states with respect to the number of shots, obtained with VQE on Callisto for two active spaces of $2$ electrons and $2$ orbitals. The red line shows the maximum standard deviation achieved for 1000 shots, while the green line indicates the same for 100000 shots. 
}}
\label{fig:conv_sim}
\end{figure*}

We also performed a convergence study of the noisy simulations on Callisto for the active space of 2 electrons and 2 orbitals. In this case we applied the noise mitigation scheme described in the main text. Fig \ref{fig:conv_sim} shows the absolute energies of R, TS and P1 states with respect to the number of shots. As expected, increasing the number of shots in the emulation reduces the standard deviation in all molecules considered. The largest reductions tend to happen around the 30000 shot value, while further shots have a smaller impact on the standard deviation.

\newpage
\bibliographystyle{apsrev4-1}
%\bibliography{references,Pluton_biblio}
\end{document}